\documentclass[letterpaper]{article} 
\usepackage{aaai2026}  
\nocopyright  
\usepackage{times}  
\usepackage{helvet}  
\usepackage{courier}  
\usepackage[hyphens]{url}  
\usepackage{graphicx} 
\usepackage{natbib}  
\usepackage{caption} 
\usepackage{algorithm}
\usepackage{algorithmic}

\usepackage{amsmath}
\usepackage{amssymb}
\usepackage{amsthm}
\usepackage{bm}

\usepackage{booktabs}
\usepackage{multirow}
\usepackage{threeparttable}

\newtheorem{proposition}{Proposition}

\title{When Is a Conformal Guarantee Fair? Auditing Silent Subgroup\\ Under-Coverage in Alzheimer's Disease Longitudinal Prediction}

\author{
    Lujia Zhong\textsuperscript{\rm 1,\rm 2},
    Xinkai Wang\textsuperscript{\rm 1,\rm 2},
    Shuo Huang\textsuperscript{\rm 1,\rm 3},
    Yonggang Shi\textsuperscript{\rm 1,\rm 2,\rm 3}
}
\affiliations{
    \textsuperscript{\rm 1}Stevens Neuroimaging and Informatics Institute, Keck School of Medicine,\\
    University of Southern California\\
    \textsuperscript{\rm 2}Ming Hsieh Department of Electrical and Computer Engineering, Viterbi School of Engineering,\\
    University of Southern California\\
    \textsuperscript{\rm 3}Alfred E. Mann Department of Biomedical Engineering, Viterbi School of Engineering,\\
    University of Southern California
}

\begin{document}
\maketitle

\begin{abstract}
Longitudinal prediction of Alzheimer's disease biomarkers increasingly informs clinical decisions, and a forecast is only useful if it also reports how much to trust it. Conformal prediction supplies this by wrapping any forecaster in a prediction band with a finite-sample coverage guarantee under exchangeability. However, standard population-level conformal prediction guarantees only marginal coverage and may mask substantial under-coverage within clinically important subgroups. We introduce a general mechanism-driven framework for auditing and repairing such subgroup under-coverage. Across two cohorts (ADNI, OASIS-3), two base forecasters, and nine attributes spanning genetic risk, demographics, and clinical severity, we find that population-level bands under-cover high-risk subgroups in 57 of 68 audited combinations, despite achieving nominal marginal coverage. We trace these failures to two mechanisms: (A) \emph{rarity}, where a group-conditional band calibrated on only $n$ patients covers at most $k/(n+1)$; and (B) \emph{tail-heaviness}, where a population-wide band is too narrow for a heavy-tailed subgroup and additional data cannot close the gap. Under-coverage falls disproportionately on patients with high genetic risk and disease severity (6.1 pp mean deficit, 95\% CI [3.3, 8.9]), while demographic groups remain at the target level on average (0.0 pp, CI [$-1.9$, 1.7]). We pair each mechanism with a corresponding conformal correction: cross-conformal pooling for rarity, per-subgroup calibration for tail-heaviness, and a coverage-safe marginal floor when both arise. Together, these corrections restore target coverage for nearly every high-risk subgroup across both cohorts and forecasters.
\end{abstract}

\section{Introduction}
\label{sec:intro}

Alzheimer's disease (AD) develops over many years, tracked through biomarkers for amyloid (A) and tau (T) pathology, neurodegeneration (N), and cognitive decline (C) \citep{jack2018nia}.
Because these markers shift before symptoms appear, forecasting future values from early visits supports trial enrichment, monitoring, and intervention timing \citep{jack2018nia,jung2025disclose}.
Traditional approaches such as linear mixed-effects models \citep{laird1982random} and multivariate progression models \citep{donohue2014longterm} capture population-level trends but struggle with nonlinear, multivariate individual dynamics \citep{oxtoby2024datadriven}.
Deep learning models, including latent ODEs \citep{rubanova2019latent,lachinov2023neuralode,wen2025lnode} and deep-kernel Gaussian processes \citep{tassopoulou2022dkgp,liang2024icdkgp}, learn these dynamics from irregularly sampled data, improving individual-level prediction.

However, even when these models predict future outcomes accurately on average, clinicians need reliable uncertainty estimates for individual predictions, especially in AD, where heterogeneous trajectories and sparse, irregular observations increase forecasting uncertainty \citep{ovadia2019trust,tassopoulou2025conformal}. Conformal prediction (CP) addresses this need by providing model-agnostic prediction bands with finite-sample coverage guarantees under exchangeability, supporting more reliable trial selection, monitoring, and intervention decisions \citep{vovk2005algorithmic,dheur2025unified}.

Standard population-level CP pools calibration residuals across all individuals to guarantee marginal coverage of at least $1-\alpha$. However, this guarantee holds only on average over the population and can mask substantial under-coverage within clinically important subgroups \citep{vovk2012conditional,barber2020limits}. Subgroup-conditional methods address this limitation by calibrating separately within prespecified groups. In particular, Mondrian CP partitions individuals by relevant attributes and provides coverage guarantees within each subgroup \citep{vovk2003mondrian,romano2020malice,jung2023batch}.

Recently, \citet{tassopoulou2025conformal} extended CP to randomly-timed AD biomarker trajectories observed at irregular clinical visits. They also introduced group-conditional bands to account for heterogeneity across demographic and clinically relevant subgroups. However, prior work lacks a more general diagnostic framework that attributes subgroup under-coverage to either finite-sample scarcity or residual-tail mismatch and pairs each failure mechanism with an appropriate conformal correction.

To fill this gap, we develop a mechanism-driven framework for auditing and repairing subgroup coverage in joint conformal forecasting. We evaluate two base forecasters across ADNI and OASIS-3 using nine attributes spanning genetic risk, demographics, and clinical severity. Despite nominal marginal coverage, population-level bands under-cover high-risk subgroups in 57 of 68 audited combinations. We trace these failures to two mechanisms, \emph{rarity} and \emph{tail-heaviness}. We then derive a corresponding repair for each: cross-conformal pooling for rarity, per-subgroup calibration for tail-heaviness, and a coverage-safe marginal floor for cells affected by both.

The main contributions are:
\begin{enumerate}
\item \textbf{Subgroup-coverage audit.} Across 17 (cohort, attribute) combinations, two base models, and two coverage levels, we show that marginal validity can mask substantial under-coverage of high-risk subgroups.

\item \textbf{Two-mechanism characterization.} We formalize two causes: \emph{rarity}, which limits small-cell coverage to $k/(n{+}1)$, and \emph{tail-heaviness}, which causes persistent under-coverage under population calibration. We validate both mechanisms empirically and show that they worsen at opposite tolerance levels.

\item \textbf{Mechanism-matched repair.} We combine cross-conformal pooling for rarity, per-subgroup calibration for tail-heaviness, and a marginal floor when both arise. The repair is model-agnostic, requires no retraining, and preserves marginal coverage.

\item \textbf{Clinical relevance.} Our method reduces coverage disparities across clinical-risk and demographic groups while providing usable prediction widths for clinician-selected biomarkers.
\end{enumerate}

\section{Related Work}
\label{sec:related}

\paragraph{Conformal prediction and conditional coverage.}
Split conformal prediction \citep{vovk2005algorithmic} provides finite-sample marginal coverage; \citet{romano2019conformalized} and \citet{lei2018distribution} refine the score via quantile regression and scale normalization.
Achieving conditional coverage uniformly over all distributions is known to be infeasible without additional assumptions \citep{vovk2012conditional,barber2020limits}.
Mondrian conformal prediction \citep{vovk2003mondrian} calibrates separately within each subgroup, but a finite-width band requires each subgroup's calibration set to be large enough ($n_a \ge \lceil 1/\alpha\rceil{-}1$, e.g.\ $19$ at $\alpha{=}.05$); with few points per cell the per-subgroup quantile becomes unreliable \citep{ding2023classcond}.

\paragraph{Fairness of prediction sets.}
\citet{romano2020malice} and \citet{jung2023batch} equalize conformal coverage across protected groups such as race and sex.
\citet{bastani2022practical} and \citet{jung2023batch} build multicalibrated predictors covering many subgroups at once.
Conditional-conformal methods \citep{gibbs2023conformal} fit coverage continuously over a covariate class via a per-group convex program.
These methods focus on protected-attribute groups that are typically well-represented in the calibration set.

\paragraph{Multi-output conformal and pooling.}
\citet{dheur2025unified} unify several multi-output conformal recipes, including M-CP (max-of-quantiles).
Cross-conformal prediction \citep{vovk2015cv} enlarges the calibration set by pooling out-of-fold scores; CV$+$ and the jackknife$+$ \citep{barber2021jackknife} give this construction a finite-sample guarantee.
\citet{guan2023localized} and \citet{hore2025local} interpolate the conformal quantile continuously across covariates.
\citet{gibbs2021adaptive} update the tolerance level online from observed coverage errors.
\citet{tassopoulou2025conformal} defines joint coverage scores by taking the max over time points; \citet{messoudi2022copula} and \citet{sun2024copulacpts} model the dependence across targets or time steps with a copula.

\paragraph{AD biomarker forecasting.}
Longitudinal prediction models for AD biomarkers include neural ODEs \citep{rubanova2019latent,lachinov2023neuralode,jung2025disclose,wen2025lnode,park2025brainode}, deep-kernel GPs \citep{tassopoulou2022dkgp,liang2024icdkgp}, and conditional normalizing flows \citep{yalavarthi2025profiti}.
We use a latent ODE and a conditional Real~NVP joint flow as the two base models, chosen because they represent opposite ends of the uncertainty spectrum; the latent ODE samples trajectories under a fixed observation-noise scale, while the flow learns a full joint density. Our calibration recipe is a post-hoc wrapper and does not depend on the choice of base model.

\section{Preliminaries and Audit Protocol}
\label{sec:prelim}

\paragraph{Problem setting.}
A patient is observed at a single baseline visit with measurements across $d$ biomarker channels (amyloid, tau, neurodegeneration, and cognition markers).
The task is to forecast the patient's future biomarker values at follow-up horizons $h\in\{12,24,48\}$~months and wrap each forecast in a prediction band that contains the true values with probability at least $1{-}\alpha$.
A base forecaster produces the prediction; conformal prediction calibrates the band around it from held-out residuals, with no distributional assumptions.

\paragraph{Split and Mondrian conformal.}
Given exchangeable calibration scores $\{s_i\}_{i=1}^n$, the split-conformal quantile at level $1-\alpha$ is the order statistic $\hat q=s_{(k)}$ with $k=\lceil(n{+}1)(1-\alpha)\rceil$, giving marginal coverage $\ge 1-\alpha$.
The textbook convention sets the band to infinite width when $\lceil(n{+}1)(1-\alpha)\rceil>n$. An infinite band is useless at the bedside, so throughout this work we instead cap the index at $k=\min(\lceil(n{+}1)(1-\alpha)\rceil,\,n)$, and a small cell returns the largest observed score rather than an infinite band.
Mondrian conformal \citep{vovk2003mondrian} partitions calibration into per-subgroup cells and uses a cell-specific $\hat q_a$; coverage holds only when the cell is large enough.

\paragraph{Joint coverage and the score.}
At horizon $h$, let $\mathcal C_h$ be the audited channels.
Because the $d$ channels have different units and scales, we normalize each per-channel residual $s_c$ by a channel-specific scale $\sigma_c$ derived from the base model's own predictive interval:
$\sigma_c=\max\!\big((\hat q_c^{\mathrm{hi}}-\hat q_c^{\mathrm{lo}})/(2z_{1-\alpha/2}),\;\kappa\,s_c^{\mathrm{res}}\big)$, $\tilde s_c=s_c/\sigma_c$.
The floor at $\kappa{=}0.25$ times the robust residual scale $s_c^{\mathrm{res}}$ (median absolute deviation of $\hat m_c{-}y_c$; insensitive over $\kappa\in[0.1,0.5]$) prevents near-zero-width intervals from dominating the max.
No held-out normalizer is needed; a point-only base model can use the training per-channel spread instead.
The joint score is the max over a visit's retained channels, $\tilde s^{\mathrm{joint}}=\max_{c}\tilde s_c$, and a visit is jointly covered when all retained channels are covered.

\paragraph{Cohorts, base models, attributes.}
We use the 400-subject forecastable ATN$+$C \textbf{ADNI} cohort \citep{petersen2010adni} (26 channels: 10~A $+$ 10~T $+$ 4~N $+$ 2~C) and, for external validation, a 658-subject 7-channel \textbf{OASIS-3} cohort \citep{lamontagne2019oasis3}.
The baseline visit is the only history input; future visits bin into $h\in\{12,24,48\}$~month.
We use 10 subject-disjoint folds (Fit $\approx$290, Cal $\approx$73, Test $\approx$37); no subject appears in two splits.
Two base models: \textbf{(B1)} the latent ODE of \citet{rubanova2019latent} and \textbf{(B2)} a conditional Real~NVP joint flow \citep{dinh2017realnvp}, both trained from scratch per fold.
Cohort characteristics, full hyperparameters, computing hardware, software versions, and random-seed handling are detailed in the technical supplement.
We audit coverage over \textbf{nine attributes}: APOE carrier and dose, sex, age, education, race, diagnosis (CN/MCI/AD), and CDR-SB and MMSE severity; on OASIS-3, all but diagnosis (which OASIS-3 stages by CDR).
For each of the eight attributes with an a-priori risk direction we fix the highest-risk subgroup \emph{before} looking at coverage (APOE4$+$, e4/e4, AD, CDR-SB$>$4, MMSE$<$24, age$\geq$80, $\leq$12\,y education, Non-White); sex has no direction, so we fix Male by convention.
Unless stated otherwise, coverage is count-pooled over horizons.

\section{The Audit: Marginal Validity Hides Broad Subgroup Under-Coverage}
\label{sec:audit}

\begin{table*}[t]
\centering
\caption{\textbf{High-risk subgroup coverage audit.} For each attribute, we fix its a priori highest-risk subgroup and report that subgroup's joint coverage (count-pooled over $h\in\{12,24,48\}$~month, JointFlow; $n$ = subgroup test visits; \emph{subscript} = width relative to pooled-M-CP). \emph{M-CP} = marginal, \emph{Mondrian} = per-subgroup, both per-fold,
no pooling, no floor (standard practice); \emph{Recipe} = per-subgroup $+$ pooling $+$ floor. \underline{Underlined} = silently under-covers ($>$1pp below nominal).}
\label{tab:breadth}
\footnotesize
\setlength{\tabcolsep}{3pt}
\begin{tabular}{ll r ccc c ccc}
\toprule
 & & & \multicolumn{3}{c}{$\alpha{=}.10$ (nom.\ .90)} & & \multicolumn{3}{c}{$\alpha{=}.05$ (nom.\ .95)} \\
\cmidrule(lr){4-6}\cmidrule(lr){8-10}
Attribute & High-risk subgroup & $n$ & M-CP & Mond. & Recipe & & M-CP & Mond. & Recipe \\
\midrule
\multicolumn{10}{@{}l}{\emph{ADNI}}\\
\quad APOE4 carrier & APOE4$+$ & 261 & \underline{0.889}\,$_{1.1}$ & 0.916\,$_{1.2}$ & \textbf{0.908}\,$_{1.2}$ & & 0.954\,$_{1.2}$ & \underline{0.931}\,$_{1.1}$ & \textbf{0.954}\,$_{1.2}$ \\
\quad APOE4 dose & e4/e4 & 45 & \underline{0.867}\,$_{1.1}$ & \underline{0.556}\,$_{0.8}$ & \textbf{0.911}\,$_{1.6}$ & & 0.956\,$_{1.2}$ & \underline{0.578}\,$_{0.7}$ & \textbf{\underline{0.933}}\,$_{1.6}$ \\
\quad Sex & Male & 277 & 0.906\,$_{1.1}$ & 0.924\,$_{1.0}$ & \textbf{0.910}\,$_{1.0}$ & & 0.978\,$_{1.2}$ & \underline{0.935}\,$_{0.9}$ & \textbf{0.960}\,$_{1.0}$ \\
\quad Age & $\geq$80 & 103 & \underline{0.854}\,$_{1.1}$ & 0.893\,$_{1.2}$ & \textbf{0.932}\,$_{1.4}$ & & 0.971\,$_{1.2}$ & \underline{0.922}\,$_{0.9}$ & \textbf{0.961}\,$_{1.2}$ \\
\quad Education & $\leq$12y & 56 & 0.911\,$_{1.1}$ & \underline{0.839}\,$_{0.9}$ & \textbf{0.929}\,$_{1.2}$ & & 0.946\,$_{1.2}$ & \underline{0.821}\,$_{0.8}$ & \textbf{0.964}\,$_{1.2}$ \\
\quad Race & Non-White & 81 & 0.951\,$_{1.1}$ & \underline{0.815}\,$_{0.8}$ & \textbf{0.951}\,$_{1.0}$ & & 1.000\,$_{1.2}$ & \underline{0.827}\,$_{0.7}$ & \textbf{0.975}\,$_{1.1}$ \\
\quad Diagnosis & AD & 55 & 0.891\,$_{1.1}$ & \underline{0.709}\,$_{1.0}$ & \textbf{0.909}\,$_{1.2}$ & & 0.945\,$_{1.1}$ & \underline{0.764}\,$_{0.9}$ & \textbf{0.945}\,$_{1.5}$ \\
\quad CDR-SB severity & CDR-SB $>$4 & 40 & \underline{0.825}\,$_{1.0}$ & \underline{0.775}\,$_{1.0}$ & \textbf{0.900}\,$_{2.1}$ & & \underline{0.875}\,$_{1.1}$ & \underline{0.725}\,$_{0.8}$ & \textbf{0.950}\,$_{1.6}$ \\
\quad MMSE severity & MMSE $<$24 & 45 & \underline{0.889}\,$_{1.1}$ & \underline{0.711}\,$_{1.0}$ & \textbf{\underline{0.889}}\,$_{1.6}$ & & 0.956\,$_{1.2}$ & \underline{0.667}\,$_{0.7}$ & \textbf{0.956}\,$_{1.6}$ \\
\midrule
\multicolumn{10}{@{}l}{\emph{OASIS-3}}\\
\quad APOE4 carrier & APOE4$+$ & 615 & 0.891\,$_{1.1}$ & 0.891\,$_{1.2}$ & \textbf{0.909}\,$_{1.0}$ & & 0.953\,$_{1.2}$ & 0.961\,$_{1.6}$ & \textbf{0.956}\,$_{1.2}$ \\
\quad APOE4 dose & e4/e4 & 89 & \underline{0.865}\,$_{1.1}$ & \underline{0.854}\,$_{1.5}$ & \textbf{0.910}\,$_{1.9}$ & & \underline{0.910}\,$_{1.2}$ & \underline{0.843}\,$_{0.9}$ & \textbf{\underline{0.933}}\,$_{2.0}$ \\
\quad Sex & Male & 695 & \underline{0.876}\,$_{1.1}$ & 0.915\,$_{1.3}$ & \textbf{0.902}\,$_{1.1}$ & & 0.942\,$_{1.2}$ & 0.965\,$_{1.6}$ & \textbf{0.950}\,$_{1.2}$ \\
\quad Age & $\geq$80 & 145 & \underline{0.848}\,$_{1.1}$ & \underline{0.834}\,$_{1.2}$ & \textbf{0.917}\,$_{1.7}$ & & \underline{0.938}\,$_{1.2}$ & \underline{0.841}\,$_{0.7}$ & \textbf{0.966}\,$_{2.3}$ \\
\quad Education & $\leq$12y & 226 & \underline{0.858}\,$_{1.1}$ & 0.907\,$_{1.6}$ & \textbf{0.912}\,$_{1.1}$ & & \underline{0.938}\,$_{1.2}$ & \underline{0.916}\,$_{1.0}$ & \textbf{0.956}\,$_{1.5}$ \\
\quad Race & Non-White & 236 & 0.907\,$_{1.1}$ & 0.936\,$_{1.6}$ & \textbf{0.924}\,$_{1.0}$ & & 0.953\,$_{1.2}$ & 0.945\,$_{1.1}$ & \textbf{0.966}\,$_{1.1}$ \\
\quad CDR-SB severity & CDR-SB $>$4 & 70 & 0.929\,$_{1.1}$ & \underline{0.771}\,$_{0.2}$ & \textbf{0.929}\,$_{1.0}$ & & 0.943\,$_{1.2}$ & \underline{0.786}\,$_{0.2}$ & \textbf{0.957}\,$_{2.5}$ \\
\quad MMSE severity & MMSE $<$24 & 81 & \underline{0.877}\,$_{1.1}$ & \underline{0.753}\,$_{0.3}$ & \textbf{0.926}\,$_{1.7}$ & & \underline{0.926}\,$_{1.2}$ & \underline{0.728}\,$_{0.2}$ & \textbf{0.963}\,$_{2.6}$ \\
\bottomrule
\end{tabular}
\end{table*}

Table~\ref{tab:breadth} reports each attribute's a priori highest-risk subgroup under three recipes: the marginal band (M-CP), per-subgroup Mondrian, and our repair (\S\ref{sec:repair}). We find that a standard recipe silently under-covers the high-risk subgroup in \textbf{57} of the \textbf{68} audited (subgroup, model, $\alpha$) combinations, which span \textbf{17} (cohort, attribute) axes, while marginal coverage holds at nominal throughout. The deficit is invisible to a marginal check, because the marginal averages over well- and poorly-covered subgroups alike.
For instance, at $\alpha{=}.05$, per-fold Mondrian covers APOE homozygotes at only $0.58$, MMSE$<$24 at $0.67$, and CDR-SB$>$4 at $0.72$ (cluster-bootstrap $95\%$ CIs $[0.51,0.78]$, $[0.54,0.84]$, $[0.60,0.84]$). At $\alpha{=}.10$, the marginal band covers CDR-SB$>$4 at $0.83$ and MMSE$<$24 at $0.89$.

\begin{figure}[t]
  \centering
  \includegraphics[width=1\columnwidth]{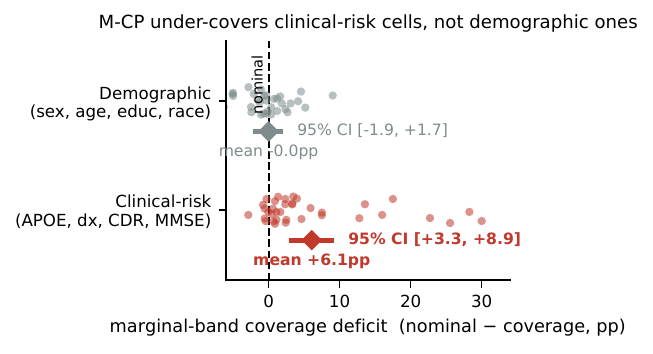}
  \caption{Marginal-band under-coverage concentrates more on \emph{clinical}-risk subgroups, not demographic ones. Each dot is one audited high-risk subgroup (deficit $=$ nominal $-$ M-CP coverage; right of zero $=$ under-covers; both cohorts and base models); diamonds are group means with $95\%$ bootstrap CI over the (cohort, attribute) axes.}
  \label{fig:clindef}
\end{figure}

We find that the deficit concentrates on high-risk subgroups while common ones remain well-covered. On ADNI (Table~\ref{tab:breadth}), the marginal band covers common demographic subgroups at nominal (e.g., Male, $n{=}277$, at $0.906/0.978$) yet under-covers clinical-risk subgroups. Across the audited (cohort, attribute) axes, clinical-risk subgroups are under-covered by a mean $6.1$~pp (95\% CI $[3.3,8.9]$, Cohen's $d{=}1.3$, $p{=}0.009$), while demographic subgroups (sex, age, education, race) sit at nominal on average ($0.0$~pp, CI $[-1.9,1.7]$). Figure~\ref{fig:clindef} visualizes this contrast: each dot is one audited high-risk subgroup, and the group means confirm that the deficit concentrates on clinical-risk axes. The demographic average is zero because individual deficits cancel out: age$\geq$80 under-covers by $\sim$5~pp and low education by up to $9$~pp, but these are offset by the well-covered majority within each axis. On the clinical axes, the under-coverage is common across subgroups. We confirm this pattern with a second base model (Latent-ODE; see supplementary material), showing that the under-coverage is consistent across both forecasters and follows clinical risk, appearing where a subgroup is heavy-tailed, rare, or both.

\section{Two Mechanisms of Silent Under-Coverage}
\label{sec:mech}

\begin{figure*}[t]
  \centering
  \includegraphics[width=0.8\textwidth]{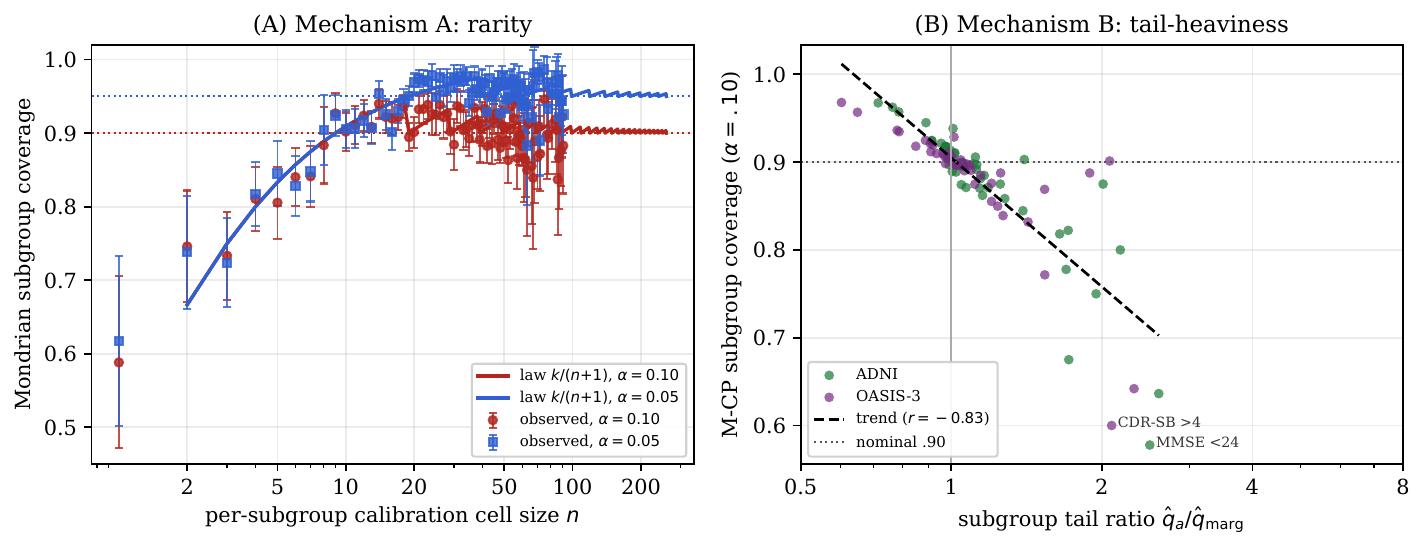}
  \caption{The two under-coverage mechanisms. \textbf{(A)~Rarity.} Per-subgroup (Mondrian) coverage vs.\ calibration cell size $n$, pooled over all attributes/cohorts/models/horizons. \textbf{(B)~Tail.} Marginal-band (M-CP) subgroup coverage falls as the subgroup's residual tail grows heavier than the population's.}
  \label{fig:mech}
\end{figure*}

We trace the under-coverage in Table~\ref{tab:breadth} to two mechanisms. To identify them, we examine how subgroup coverage depends on two quantities: the number of calibration patients in a subgroup's cell, and the heaviness of the subgroup's residual tail relative to the population. Figure~\ref{fig:mech} separates these two effects: Panel~A shows that per-subgroup (Mondrian) coverage drops when the calibration cell is small, and Panel~B shows that marginal-band (M-CP) coverage drops when the subgroup's residual tail is heavier than the population's. Each is the failure mode of a different standard conformal method. We formalize each below and validate it against the empirical coverage in Figure~\ref{fig:mech}.

\paragraph{Mechanism A: rarity.}
Per-subgroup (Mondrian) conditioning gives each stratum its own quantile, but that
quantile is estimated on a small cell. This yields a closed-form deficit.

\begin{proposition}[Small-cell coverage law]
\label{prop:rarity}
Let a Mondrian cell hold $n\ge1$ exchangeable calibration scores and calibrate with
$k=\min(\lceil(n{+}1)(1-\alpha)\rceil,\,n)$. A fresh exchangeable test point from the
same cell is covered with probability $k/(n{+}1)$. In particular, once
$n<\lceil 1/\alpha\rceil-1$ the cell is \emph{starved}: $k=n$ and the coverage is
exactly $n/(n{+}1)<1-\alpha$.
\end{proposition}

Proposition~\ref{prop:rarity} restates the standard finite-sample conformal coverage identity \citep{vovk2005algorithmic,lei2018distribution,angelopoulos2021gentle} in the starved-cell regime. When $\lceil(n{+}1)(1-\alpha)\rceil>n$, the order statistic that split conformal asks for does not exist among the $n$ calibration scores, and the usual remedy is to set the band to infinite width. An infinite band contains every value and tells a clinician nothing, so our engines use the largest calibration score instead, and the $n/(n{+}1)$ coverage above follows from that substitution. Fig.~\ref{fig:mech}A pools every Mondrian cell across all attributes, cohorts, models, and horizons. Observed coverage tracks $k/(n{+}1)$ with negligible bias (ADNI-JointFlow bias $-0.002$ at $\alpha{=}.05$ over 1346 cells; median $|\mathrm{bias}|$ across all cohort/model/$\alpha$ is $0.002$) and falls below nominal for small $n$, reaching $\sim0.60$ at $n{=}1$. The proposition predicts how much a given subgroup under-covers and the minimum cell size needed to fix it. Rarity is worst at $\alpha{=}.05$, where the starvation threshold $\lceil 1/\alpha\rceil-1$ is highest.

\paragraph{Mechanism B: tail-heaviness.}
The marginal band (M-CP) uses the entire cohort as its calibration cell, so it does not suffer from small samples. Instead, a single quantile is too narrow for a subgroup whose residual tail is heavier than the population's. This is a distributional deficit that persists regardless of sample size, complementing the finite-sample rarity of Proposition~\ref{prop:rarity}.

\begin{proposition}[Tail-driven marginal deficit]
\label{prop:tail}
Work in the population limit. Let $F$ be the pooled score CDF and $F_a$ that of subgroup $a$, let the marginal band use $q_{\mathrm{marg}}{=}F^{-1}(1{-}\alpha)$, and let $q_a{:=}\inf\{t:F_a(t)\ge1{-}\alpha\}$. Subgroup $a$'s coverage under this band is exactly $F_a(q_{\mathrm{marg}})$, so $a$ is under-covered iff its own quantile exceeds the marginal, $q_a{>}q_{\mathrm{marg}}$. In that case, if $F_a$ has a non-increasing upper-tail density $f_a$ on $[q_{\mathrm{marg}},q_a]$, the deficit satisfies
\begin{equation*}
\begin{aligned}
  (1{-}\alpha)-F_a(q_{\mathrm{marg}})
  &=\textstyle\int_{q_{\mathrm{marg}}}^{q_a} f_a\\[-1pt]
  &\in\big[f_a(q_a)\,\Delta,\;f_a(q_{\mathrm{marg}})\,\Delta\big],
\end{aligned}
\end{equation*}
with $\Delta{:=}q_a{-}q_{\mathrm{marg}}$. At fixed $q_{\mathrm{marg}}$ and $f_a$ the deficit is non-decreasing in the tail gap $\Delta$ (equivalently in the ratio $q_a/q_{\mathrm{marg}}$) and \emph{does not depend on calibration size}. An oracle subgroup band uses $q_a$ in place of $q_{\mathrm{marg}}$ and removes the deficit exactly. A finite-sample subgroup band estimates $q_a$ from the cell, so it removes this deficit and incurs the small-cell deficit of Prop.~\ref{prop:rarity} in its place.
\end{proposition}

The two propositions form a matched pair (proof of Prop.~\ref{prop:tail} in the technical supplement). Rarity is a finite-sample artifact that vanishes as $n\!\to\!\infty$. The tail deficit is a distributional mismatch that persists at any sample size, and in the population limit only conditioning removes it. The supplement also bounds the tail deficit by $\alpha(1{-}\pi_a)/\pi_a$ for a subgroup with population share $\pi_a$, so a subgroup that holds a large share of the cohort cannot show a large tail deficit. In summary, both mechanisms are worse for smaller subgroups: Prop.~\ref{prop:rarity} says a subgroup under-covers when its calibration cell has few patients, and this bound says only a subgroup that is a small fraction of the cohort can show a large tail deficit. Fig.~\ref{fig:mech}B confirms the tail mechanism: among large cells, M-CP subgroup coverage falls with the tail ratio $\hat q_a/\hat q_{\mathrm{marg}}$ as Prop.~\ref{prop:tail} predicts ($r{=}{-}0.83$ over 90 cells across both base models; attribute-block bootstrap 95\% CI $[{-}0.91,{-}0.75]$). The heaviest-tailed clinical subgroups (CDR-SB$>$4, MMSE$<$24) are the most under-covered.

\paragraph{The dichotomy.}
The two mechanisms are the failure modes of the two standard methods, and they worsen at \emph{opposite} tolerances: rarity (Mondrian) worsens at $\alpha{=}.05$, where the starvation threshold is higher; the tail deficit (M-CP) worsens at $\alpha{=}.10$, where the marginal band is tighter.

\section{Mechanism-matched Repair}
\label{sec:repair}

We address each mechanism (Props.~\ref{prop:rarity},~\ref{prop:tail}) with a corresponding conformal operation: \textbf{conditioning} removes the tail (B), \textbf{cross-conformal pooling} addresses the rarity (A), and a \textbf{marginal floor} handles the residual where both act at once (subgroup intersections, \S\ref{sec:inter}). For a sensitive attribute $A$ with groups $\{a\}$, we apply the three ingredients in sequence. The order matters: conditioning creates the rarity that pooling then removes.

\paragraph{(1)~Conditioning.} A per-stratum quantile $\hat q_a$ gives each heavy-tailed subgroup its own, correctly wider band, removing mechanism~B. This is equivalent to Mondrian conditioning by attribute $A$.

\paragraph{(2)~Cross-conformal pooling.} Conditioning reduces the cell size and creates mechanism~A. Leakage-safe cross-conformal pooling \citep{vovk2015cv,barber2021jackknife} restores it. For test fold $k$, the calibration set is the union of the other nine folds' out-of-fold scores, $\mathcal C_k^{\mathrm{pool}}=\bigcup_{j\neq k}\{s_i:i\in\mathrm{Test}_j\}$, each scored by a model that did not train on subject $i$. This lifts each per-(stratum, horizon) cell from $\sim$17 to $\sim$78 joint visits, above the finite-sample cell-size threshold at both tolerances. With enough calibration points, $\hat q_a$ stops under-covering.

\paragraph{(3)~Marginal floor.} When even the pooled cell is thin, the stratum quantile is floored at the marginal:
\begin{equation}
\hat q_a^{*}=\max\!\big(\hat q_a,\;\hat q_{\mathrm{marg}}\big).
\label{eq:shrink}
\end{equation}
This one-sided floor only raises a quantile, never lowers one. Because $\hat q_{\mathrm{marg}}$ is the mixture quantile, a heavy-tailed high-risk subgroup has $\hat q_a>\hat q_{\mathrm{marg}}$, so the floor is \emph{inactive} for that subgroup, which keeps its own larger quantile. The floor lifts only the light-tailed or starved cells. A two-sided shrinkage would instead pull the heavy-tailed subgroup's quantile down and under-cover it. The floor needs no tuning or selection set. Because $\hat q_a^*\ge\hat q_{\mathrm{marg}}$ everywhere, the recipe's mixture coverage is at least that of pooled split~CP, whose guarantee is $1-2\alpha-O(K^{-1})$ for $K$ cross-conformal folds ($K{=}10$) \citep{vovk2015cv,barber2021jackknife}. The recipe therefore inherits marginal validity.

The full procedure is given as Algorithm~1 in the technical supplement, where the target-channel set $\mathcal{C}$ chooses which channels the joint band must cover. With all channels, we audit the full joint band.

\begin{table}[t]
\centering
\caption{\textbf{Natural partial repairs vs.\ the full recipe} (ADNI, JointFlow; joint coverage count-pooled over $h$; \emph{subscript} = width relative to pooled M-CP). We track \emph{one subgroup per mechanism} down the ingredient ladder: mechanism-B probe APOE4$+$ (heavy-tailed but common) and mechanism-A probe CDR-SB$>$4 (rare).}
\label{tab:ablation}
\footnotesize
\setlength{\tabcolsep}{2pt}
\begin{tabular}{@{}l cc c cc@{}}
\toprule
 & \multicolumn{2}{c}{B: APOE4$+$} & & \multicolumn{2}{c}{A: CDR-SB $>$4} \\
\cmidrule(lr){2-3}\cmidrule(lr){5-6}
Ingredient & $\alpha{=}.10$ & $\alpha{=}.05$ & & $\alpha{=}.10$ & $\alpha{=}.05$ \\
\midrule
M-CP & \underline{0.889}\,$_{1.1}$ & 0.954\,$_{1.2}$ & & \underline{0.825}\,$_{1.0}$ & \underline{0.875}\,$_{1.1}$ \\
\quad $+$ pooling & \underline{0.862}\,$_{1.0}$ & \underline{0.931}\,$_{1.0}$ & & \underline{0.750}\,$_{1.0}$ & \underline{0.850}\,$_{1.0}$ \\
Mondrian & 0.916\,$_{1.2}$ & \underline{0.931}\,$_{1.1}$ & & \underline{0.775}\,$_{1.0}$ & \underline{0.725}\,$_{0.8}$ \\
\quad $+$ pooling & 0.908\,$_{1.2}$ & 0.950\,$_{1.2}$ & & 0.900\,$_{2.1}$ & 0.950\,$_{1.6}$ \\
\quad $+$ floor (\textbf{Recipe}) & 0.908\,$_{1.2}$ & 0.954\,$_{1.2}$ & & 0.900\,$_{2.1}$ & 0.950\,$_{1.6}$ \\
\bottomrule
\end{tabular}
\end{table}

\paragraph{Which ingredient fixes which mechanism (Table~\ref{tab:ablation}).}
We isolate each ingredient on two probe subgroups (Table~\ref{tab:ablation}), one per mechanism: APOE4$+$ (heavy-tailed but common, testing mechanism~B) and CDR-SB$>$4 (rare, testing mechanism~A).
The marginal band under-covers APOE4$+$ at $0.889$ ($\alpha{=}.10$). Conditioning fixes that tail, raising coverage to $0.916$. However, conditioning also shrinks the calibration cell: on CDR-SB$>$4 the per-fold Mondrian cell is too small, and coverage falls to $0.725$ ($\alpha{=}.05$). Cross-conformal pooling restores that cell ($0.725{\to}0.950$). Together, conditioning and pooling cover both subgroups. Pooling the marginal band alone, without conditioning, leaves the tail deficit in place ($0.889{\to}0.862$): mechanism~B requires conditioning, not more data. The floor moves coverage by at most $0.4$~pp in this single-attribute setting (on APOE4$+$ at $\alpha{=}.05$, where pooling had already reached nominal) and becomes necessary only at subgroup intersections (\S\ref{sec:inter}). We confirm the same pattern on the Latent-ODE base model (see supplementary material).

\paragraph{Aggregate repair.}
Across both base models (JointFlow in Table~\ref{tab:breadth}; Latent-ODE in supplementary material), our recipe raises subgroup coverage over the marginal band by a mean $4.3$~pp across the $17$ audited axes (axis-level sign-flip permutation $p{=}3.5{\times}10^{-4}$; CI $[2.3,6.4]$; $88\%$ of axes improve). The gain is larger on the $9$ clinical axes: $+6.7$~pp ($p{=}3.2{\times}10^{-3}$). Over per-fold Mondrian, the improvement is $9.5$~pp ($p{=}1.5{\times}10^{-4}$). Our recipe requires no retraining; it is a post-hoc quantile computation. Its coverage sits at or above nominal on average, with clinical subgroups over-covered by $0.6$~pp.

\begin{figure}[t]
  \centering
  \includegraphics[width=0.68\columnwidth]{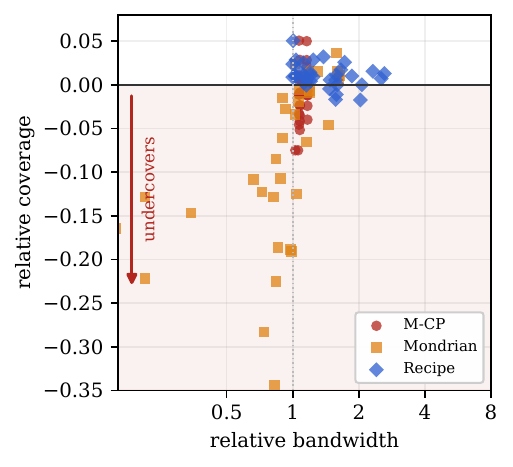}
  \caption{Coverage vs.\ width. Each point is a high-risk subgroup under one method, with x = relative bandwidth to the pooled M-CP band and y = coverage - nominal.}
  \label{fig:frontier}
\end{figure}

\paragraph{Coverage at minimal width (Fig.~\ref{fig:frontier}).}
The coverage improvement does not come from uniformly wider bands. Per-fold M-CP under-covers 44\% of high-risk subgroups at a median $1.1\times$ width. Per-fold Mondrian is narrower (median $0.95\times$) yet under-covers 71\%; on a rare cell, its band collapses to as little as $0.2\times$ while missing the subgroup (mechanism~A). Our recipe covers 91\% at median $1.2\times$, wide only where the tail demands it (e.g., $2.1\times$ for CDR-SB$>$4). Conditioning is efficient: it widens the band only for the heavy-tailed subgroup, while a marginal band wide enough for that subgroup applies the same width to everyone ($1.0\times$ for APOE4$-$ under the recipe vs.\ $1.2\times$ for all of APOE at $\alpha{=}.05$). Dropping the marginal floor is width-minimal but loses the safety net on the rarest cells.

\begin{table}[t]
\centering
\caption{\textbf{Efficiency vs.\ a SOTA conditional-conformal baseline (Gibbs--Cand\`es).} cov. = mean coverage over each block's high-risk subgroups; wid = median width relative to pooled M-CP, aggregated over both cohorts.}
\label{tab:gibbs}
\footnotesize
\setlength{\tabcolsep}{4pt}
\begin{tabular}{@{}ll cc c cc@{}}
\toprule
 & & \multicolumn{2}{c}{Gibbs--Cand\`es} & & \multicolumn{2}{c}{Recipe} \\
\cmidrule(lr){3-4}\cmidrule(lr){6-7}
Base model & Cells & cov. & wid & & cov. & wid \\
\midrule
\multicolumn{7}{@{}l}{\emph{$\alpha{=}.10$ (nominal~.90)}}\\
\quad JointFlow & clinical-risk & 0.916 & 1.23 & & 0.913 & \textbf{1.17} \\
\quad JointFlow & demographic & 0.909 & 1.05 & & 0.906 & \textbf{1.03} \\
\quad Latent-ODE & clinical-risk & 0.915 & 2.35 & & 0.898 & \textbf{1.48} \\
\quad Latent-ODE & demographic & 0.913 & \textbf{0.99} & & 0.906 & 1.05 \\
\midrule
\multicolumn{7}{@{}l}{\emph{$\alpha{=}.05$ (nominal~.95)}}\\
\quad JointFlow & clinical-risk & 0.957 & 1.68 & & 0.955 & \textbf{1.42} \\
\quad JointFlow & demographic & 0.958 & \textbf{1.02} & & 0.958 & 1.06 \\
\quad Latent-ODE & clinical-risk & 0.968 & 1.87 & & 0.957 & \textbf{1.44} \\
\quad Latent-ODE & demographic & 0.959 & \textbf{1.00} & & 0.957 & 1.13 \\
\bottomrule
\end{tabular}
\end{table}

\paragraph{A SOTA conditional-conformal baseline.}
A natural alternative is to fit coverage continuously over the subgroup space. Table~\ref{tab:gibbs} compares the Gibbs--Cand\`es conditional-conformal estimator \citep{gibbs2023conformal} ($\Phi$ the subgroup one-hot), evaluated on identical scores, calibration, and test visits. The comparison corroborates our mechanism diagnosis: conditioning alone recovers the subgroups, matching our coverage everywhere (mean $|\Delta\text{cov}|{=}0.01$ over $68$ combinations). This confirms that conditioning closes the tail deficit (Prop.~\ref{prop:tail}), independent of cell size or score choice. However, the Gibbs--Cand\`es estimator produces wider bands on the rare heavy-tailed clinical subgroups ($1.68\times$ vs.\ our $1.42\times$ on JointFlow at $\alpha{=}.05$; up to $2.35\times$ vs.\ $1.48\times$ on Latent-ODE at $\alpha{=}.10$). On well-populated demographic subgroups, the two methods are indistinguishable. Our recipe attains the same coverage with up to $\sim$35\% less width on the clinical subgroups, in closed form (no per-group convex program), and preserves the marginal guarantee by construction across both cohorts and base models.

\section{When Rarity Meets Heavy Tails}
\label{sec:inter}

\begin{table*}[!t]
\centering
\caption{\textbf{Intersectional subgroups.} Joint coverage of high-risk intersections of two attributes (JointFlow, cross-conformally pooled; \emph{subscript} = width relative to pooled M-CP; $n$ = median calibration cell size). M-CP: marginal. Mond.\,($a_1$): condition on the primary attribute. Mond.\,($\cap$): condition on the intersection. All four methods here use pooling, showing that the pooling alone no longer suffices, and the marginal floor is required. \underline{Underline} = under-covers.}
\label{tab:intersectional}
\footnotesize
\setlength{\tabcolsep}{3pt}
\begin{tabular}{ll r cccc c cccc}
\toprule
 & & & \multicolumn{4}{c}{$\alpha{=}.10$ (nom.\ .90)} & & \multicolumn{4}{c}{$\alpha{=}.05$ (nom.\ .95)} \\
\cmidrule(lr){4-7}\cmidrule(lr){9-12}
Cohort & Intersection & $n$ & M-CP & M.\,$a_1$ & M.\,$\cap$ & Recipe & & M-CP & M.\,$a_1$ & M.\,$\cap$ & Recipe \\
\midrule
ADNI & APOE4$+$ $\wedge$ AD & 17 & \underline{0.781}\,$_{1.0}$ & \underline{0.844}\,$_{1.2}$ & \underline{0.875}\,$_{2.0}$ & \textbf{\underline{0.875}}\,$_{2.0}$ & & \underline{0.844}\,$_{1.0}$ & \underline{0.906}\,$_{1.2}$ & \underline{0.906}\,$_{1.5}$ & \textbf{\underline{0.906}}\,$_{1.5}$ \\
 & APOE4$+$ $\wedge$ CDR-SB$>$4 & 12 & \underline{0.636}\,$_{1.0}$ & \underline{0.727}\,$_{1.2}$ & \underline{0.864}\,$_{2.2}$ & \textbf{\underline{0.864}}\,$_{2.2}$ & & \underline{0.727}\,$_{1.0}$ & \underline{0.773}\,$_{1.2}$ & \underline{0.909}\,$_{1.6}$ & \textbf{\underline{0.909}}\,$_{1.6}$ \\
 & APOE4$+$ $\wedge$ age$\geq$80 & 18 & \underline{0.791}\,$_{1.0}$ & \underline{0.837}\,$_{1.2}$ & 0.907\,$_{1.3}$ & \textbf{0.907}\,$_{1.3}$ & & \underline{0.907}\,$_{1.0}$ & \underline{0.930}\,$_{1.2}$ & \underline{0.930}\,$_{1.2}$ & \textbf{\underline{0.930}}\,$_{1.2}$ \\
 & APOE4$+$ $\wedge$ MMSE$<$24 & 17 & \underline{0.714}\,$_{1.0}$ & \underline{0.821}\,$_{1.2}$ & 0.893\,$_{1.6}$ & \textbf{0.893}\,$_{1.6}$ & & \underline{0.821}\,$_{1.0}$ & \underline{0.893}\,$_{1.2}$ & \underline{0.893}\,$_{1.6}$ & \textbf{\underline{0.929}}\,$_{1.6}$ \\
 & APOE4$+$ $\wedge$ Non-White & 10 & \underline{0.865}\,$_{1.0}$ & 0.919\,$_{1.2}$ & 0.919\,$_{1.4}$ & \textbf{0.919}\,$_{1.4}$ & & 0.946\,$_{1.0}$ & 0.973\,$_{1.2}$ & \underline{0.919}\,$_{1.1}$ & \textbf{0.946}\,$_{1.1}$ \\
 & Male $\wedge$ AD & 19 & \underline{0.838}\,$_{1.0}$ & \underline{0.838}\,$_{1.0}$ & 0.919\,$_{1.5}$ & \textbf{0.919}\,$_{1.5}$ & & \underline{0.892}\,$_{1.0}$ & 0.946\,$_{1.0}$ & 0.946\,$_{1.0}$ & \textbf{0.946}\,$_{1.0}$ \\
\midrule
OASIS-3 & APOE4$+$ $\wedge$ CDR-SB$>$4 & 17 & \underline{0.886}\,$_{1.0}$ & \underline{0.886}\,$_{1.0}$ & 0.909\,$_{2.6}$ & \textbf{0.932}\,$_{2.6}$ & & \underline{0.909}\,$_{1.0}$ & \underline{0.932}\,$_{1.2}$ & \underline{0.909}\,$_{2.5}$ & \textbf{\underline{0.932}}\,$_{2.5}$ \\
 & APOE4$+$ $\wedge$ MMSE$<$24 & 24 & \underline{0.879}\,$_{1.0}$ & \underline{0.862}\,$_{1.0}$ & 0.931\,$_{1.8}$ & \textbf{0.931}\,$_{1.8}$ & & \underline{0.914}\,$_{1.0}$ & 0.948\,$_{1.2}$ & 0.948\,$_{2.5}$ & \textbf{0.948}\,$_{2.5}$ \\
 & Male $\wedge$ CDR-SB$>$4 & 14 & \underline{0.882}\,$_{1.0}$ & 0.912\,$_{1.1}$ & 0.912\,$_{3.9}$ & \textbf{0.941}\,$_{3.9}$ & & \underline{0.912}\,$_{1.0}$ & \underline{0.912}\,$_{1.2}$ & \underline{0.912}\,$_{2.5}$ & \textbf{0.941}\,$_{2.5}$ \\
\bottomrule
\end{tabular}
\end{table*}

\begin{figure}[t]
  \centering
  \includegraphics[width=1\columnwidth]{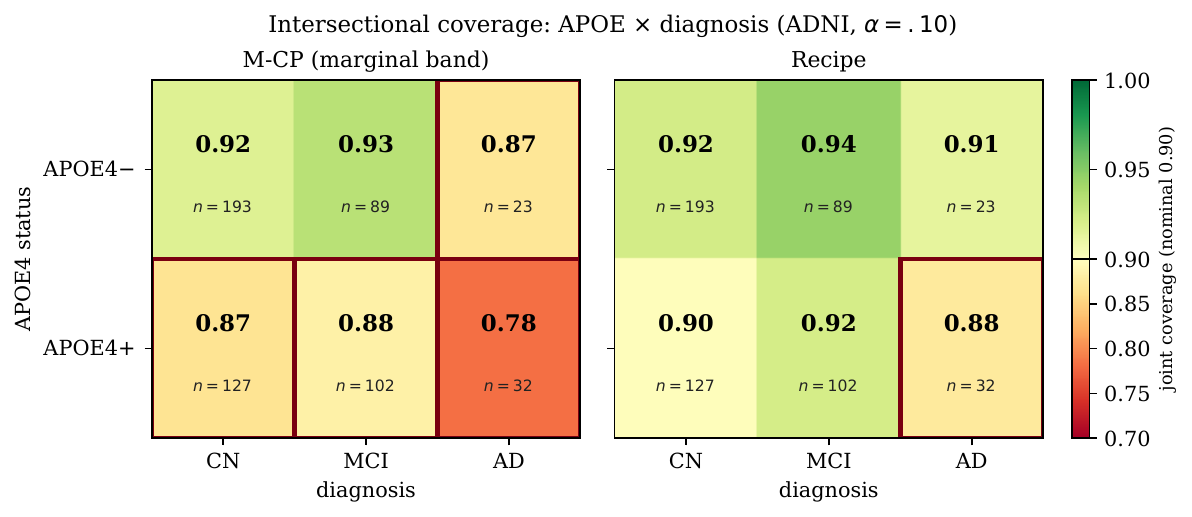}
  \caption{Intersectional coverage on APOE$\times$diagnosis (ADNI, JointFlow, $\alpha{=}.10$).}
  \label{fig:intersect}
\end{figure}

The highest-risk patients occupy the intersection of risk axes (e.g., an APOE4 carrier who also has AD), where both mechanisms act at once: the cell is heavy-tailed and rare. On the APOE$\times$diagnosis grid (Fig.~\ref{fig:intersect}), the marginal band under-covers the entire APOE4$+$ row and AD column. Our recipe lifts them toward nominal, restoring all but the rarest APOE4$+\wedge$AD corner, which marks the data-limited frontier. Table~\ref{tab:intersectional} shows the coverage of each ingredient on these intersections.

The marginal band under-covers these intersections worst of all. Even conditioning on a single attribute is often insufficient, because the second attribute's tail remains within the conditioned stratum. For example, for APOE4$+\wedge$CDR-SB$>$4 at $\alpha{=}.10$, coverage moves only from $0.64$ to $0.73$. Conditioning on the intersection itself removes the tail but leaves a cell of only $n\!\sim\!12$--$17$, too small to cover at $\alpha{=}.05$: APOE4$+\wedge$MMSE$<$24 falls to $0.89$. The marginal floor partially recovers it, lifting that same cell from $0.89$ to $0.93$.

The marginal floor, which barely moved coverage in the single-attribute setting (because pooling was already sufficient for APOE alone), becomes necessary here. However, the frontier is visible: a cell both extremely rare and heavy-tailed (APOE4$+\wedge$CDR-SB$>$4, $n\!\sim\!12$) remains near $0.91$ even under the recipe, beyond what any post-hoc quantile can fix. The complete set of measurable high-risk intersections and their replication on the Latent-ODE base model appear in the technical supplement.

\section{Discussion}
\label{sec:discussion}

\paragraph{From hidden failure to repair.}
We show that a marginal guarantee holds on average even as it fails the patients it is meant to protect. Over-coverage of the well-predicted majority and under-coverage of the hard-to-predict minority cancel out, so the overall rate reveals nothing about any single subgroup a clinician faces. We trace the under-coverage to two mechanisms that worsen at opposite tolerance levels, so no single adjustment fixes both. We give a direct fix for each: pooling addresses rarity, conditioning addresses the tail, and a coverage-safe floor handles the cases where both mechanisms overlap.

\paragraph{Beyond Alzheimer's.}
Our audit and repair do not depend on AD biomarkers specifically. Both mechanisms arise from the structure of conformal prediction itself. Because our repair is a post-hoc quantile that leaves the base forecaster untouched, we show that the same recipe transfers across two cohorts and two very different forecasters (normalizing-flow and latent-ODE). Any setting that applies marginal conformal bands to a population with known high-risk subgroups is expected to face a similar failure and can apply the same mechanism-matched fix. Alzheimer's forecasting makes the stakes concrete: the high-risk subgroups are named ahead of time, and failing to cover them has a direct clinical cost.

\paragraph{Valid and Usable.}
A joint band over all 26 channels is conservative for any single channel, because its width is determined by the least predictable channel. As a result, the band over-covers well-predicted markers at a width too wide to be clinically useful. Since we define the band over a target-channel set $\mathcal{C}$ (Algorithm~1 in the supplement), a clinician can set $\mathcal{C}{=}\{$MMSE, CDR-SB$\}$ and our recipe returns a tighter band valid jointly over those two scores (shown in supplementary material).

\paragraph{External validation and limits.}
OASIS-3 reproduces the pattern, if anything more sharply: on Latent-ODE the marginal band under-covers APOE4 carriers more than it does on ADNI, and our recipe restores coverage to nominal. Because its imaging is sparse at future visits, its evidence is strongest for cognition. Our recipe restores at least nominal coverage for every high-risk subgroup of Table~\ref{tab:breadth} except the rarest (APOE homozygotes, MMSE$<$24, $n_{\mathrm{cal}}{<}20$) and the intersections that are at once extremely rare and heavy-tailed. That residue is a data-limited frontier that post-hoc quantile cannot certify $1-\alpha$ from fewer than $\lceil1/\alpha\rceil-1$ points. Two further limits are the single-history protocol and ADNI's small minority-race subgroup ($n{=}60$).

\section{Conclusion}
\label{sec:conclusion}

We show that for multi-horizon AD longitudinal prediction, a marginal conformal guarantee can silently under-cover the high-risk subgroups, a pattern we confirm across both cohorts. We trace the under-coverage to two mechanisms, rarity and tail-heaviness. Both are known weaknesses of standard conformal methods, and we show that the two combine in the most vulnerable patients. We address each with a corresponding fix: per-subgroup calibration, cross-conformal pooling, and a coverage-safe floor. Together, these restore coverage across nine attributes and two cohorts, except for the smallest subgroups, where calibration data are too scarce for any method to provide a valid band. The result is a drop-in procedure for auditing whether a conformal guarantee is clinically fair and for repairing it when the overall average hides an under-covered subgroup.

\section*{Acknowledgment}
This work was supported by the National Institute of Health (NIH) under grants R01EB022744, RF1AG077578,  R01AG064584, U19AG078109, and P30AG066530.

Data used in preparation of this article were obtained from the Alzheimer’s Disease Neuroimaging Initiative (ADNI) database (adni.loni.usc.edu). As such, the investigators within the ADNI contributed to the design and implementation of ADNI and/or provided data but did not participate in analysis or writing of this report. A complete listing of ADNI investigators can be found at: http://adni.loni.usc.edu/wp-content/uploads/how\_to\_apply/\\ADNI\_Acknowledgement\_List.pdf

\bibliography{refs}

\end{document}